\documentclass[12pt]{article}
\usepackage{amsmath,latexsym}
\usepackage{graphicx}
\begin{document}

 \title{\Huge Gravitational lensing by stable C-field wormhole}
 \author{F.Rahaman$^*$, M.Kalam$^{\ddag}$
 and S.Chakraborty$^\dag$}

\date{}
 \maketitle
 \begin{abstract}
It has been recently shown that Hoyle-Narlikar's C-field theory
admits wormhole geometry.  We derive the deflection angle of
light rays caused by C-field wormhole in the strong field limit
approach of gravitational lensing theory. The linearized
stability of C-field wormhole under spherically symmetric
perturbations about static equilibrium is also explored.
\end{abstract}

  \footnotetext{ Pacs Nos :  04.20 Gz,04.50 + h, 04.20 Jb   
\\
 $*$Dept.of Mathematics, Jadavpur University, Kolkata-700 032, India
                                  E-Mail:farook\_rahaman@yahoo.com\\
$\ddag$Dept. of Phys. , Netaji Nagar College for Women, Regent Estate, Kolkata-700092, India.\\
  $\dag$Dept. of Maths., Meghnad Saha Institute of Technology,
                                           Kolkata-700150, India
}
    \mbox{} \hspace{.2in}

\title{\Huge Introduction: }

Wormholes are classical or quantum solutions for the gravitational
field equations describing a bridge between two asymptotic
manifolds. Classically, they can be interpreted as instantons
describing a tunneling between two distant regions. In a pioneer
work, Morris and Thorne [1] have shown that the construction of
wormhole would require a very unusual form of stress energy
tensor. The matter that characterized the above stress energy
tensor is known as exotic matter. This exotic i.e. hypothetical
matter can be of the following form either energy density of
matter $\rho < 0$ or  $\rho > 0$ but pressure $p< 0$. Till now,
we do not know where and how the exotic matter could be
collected. So, Scientists have assumed several alternative
sources such as Phantom energy [2-6], Chaplygin gas[7], Tachyon
field [8], Casimir field [9] etc for exotic matter source. Also,
some authors have used alternative theories to obtain wormhole
geometry [10-19]. Long ago, since 1966, Hoyle and Narlikar [HN]
proposed an alternative theory of gravity known as C-field theory
[20]. HN adopted a field theoretic approach introducing a massless
and chargeless scalar field C in the Einstein-Hilbert action to
account for the matter creation.

A C-field generated by a certain source equation, leads to
interesting change in the cosmological solution of Einstein field
equations. Several authors,  have studied cosmological models and
topological defects in presence of C-field [20].

Recently, Rahaman et al [21] have pointed out that a spherically
symmetric vacuum solutions to the C-field theory give rise to a
wormhole.  One of the most important applications of general
relativity is the deflection of light by a gravitating body. This
deflection of light by a gravitational field is known as
gravitational lens. In recent time, gravitational lensing from a
strong field perspective is a very active area of research. A few
years back, Virbhadra et al [22] had discovered an analytic method
to calculate the deflection for any spherically symmetric space
time in the strong field limit. Cramer et al [23] have discussed
wormhole lensing in the weak field. Recently, Tejeirio [24] and
Nandi et al [25] have studied gravitational lensing by traversable
wormhole in the strong field limit. By using the method in
Ref.[22], we calculate the lensing effect of the C-field wormhole.
A comparison of the deflection angle has been made between the
C-field wormhole solution with Schwarzschild solution. Now it
will be interesting to investigate the stability of this C-field
wormhole. To test stability, we match the interior C-field
wormhole geometry with an exterior Schwarzschild vacuum solution
at a junction interface. Assuming thin shell Schwarzschild
wormhole and using cut and paste technique[26-32], we analyze the
stability of this wormhole to linearized spherically symmetric
perturbations around static equilibrium solutions.

\title{\Huge2. C-field wormhole: }

  The modified
Einstein equations due to HN through the introduction of an
external C - field are
      \begin{equation}
              R^{ab} - \frac{1}{2}g^{ab}R
             = - 8\pi G [T^{ab}- f{ C^aC^b +
             \frac{1}{2}fg^{ab}C^iC_i}]
           \end{equation}
      where C, a scalar field representing creation of matter, $x_i$ , i = 0,1,2,3 stand for the

space-time coordinates with
\begin{equation} C_i = \frac{\partial C}{ \partial
x_i} \end{equation}  and $T_{ab}$ is the matter tensor and $f
>  0$.

Let us consider the static spherically symmetric metric as
\begin{equation}
               ds^2=  - e^\nu dt^2+ e^\mu dr^2+r^2( d\theta^2+sin^2\theta
               d\phi^2)
         \label{Eq3}
          \end{equation}

The independent field equations for the metric (3) are
\begin{equation}e^{-\mu}
[\frac{1}{r^2}-\frac{\mu^\prime}{r}]-\frac{1}{r^2}= 4\pi
Gfe^{-\mu}{(C^\prime)^2}\end{equation}
\begin{equation}e^{-\mu}
[\frac{1}{r^2}+\frac{\nu^\prime}{r}]-\frac{1}{r^2}= -4\pi
Gfe^{-\mu}{(C^\prime)^2}\end{equation}
\begin{equation}e^{-\mu}
[\frac{1}{2}(\nu^\prime)^2+ \nu^{\prime\prime}
-\frac{1}{2}\mu^\prime\nu^\prime + \frac{1}{r}({\nu^\prime-
\mu^\prime})] = 8\pi Gfe^{-\mu}{(C^\prime)^2}\end{equation}

The solutions are given by [21]

\begin{equation}e^{\nu}= constant \end{equation}
\begin{equation}e^{-\mu}= 1-\frac{D}{r^2}\end{equation}
\begin{equation}C=\frac{1}{\sqrt(4\pi
Gf)}\sec^{-1}\frac{r}{\sqrt(D)}+ C_0\end{equation}

where D and $C_0$ are integration constants.

 Thus the metric (3) can be written in Morris-Thorne
cannonical form as

\begin{equation}
               ds^2=  - e^{\nu}dt^2+ \frac{dr^2}{[ 1-\frac{b(r)}{r}]}+r^2( d\theta^2+sin^2\theta
               d\phi^2)
          \end{equation}
Here,  $ b(r) =\frac{D}{r}$  is called shape function and $e^{\nu}
= $ redshift  function $   =  constant $ .

 The
throat of the wormhole occurs at $ r = r_0 =  \sqrt{D}>0 $. One
can note  that since now $r\geq r_0 > 0$, there is no horizon.

\title{\Huge3. Gravitational Lensing: }

Now,  we match the interior wormhole solution to the exterior
Schwarzschild solution ( in the absence of C-field ). To match
the interior to the exterior, we impose only the continuity of the
metric coefficients, $ g_{\mu\nu} $, across a surface, S , i.e. $
{g_{\mu\nu}}_{(int)}|_S =  {g_{\mu\nu}}_{(ext)}|_S $.

[ This condition is not sufficient to different space times.
However, for space times with a good deal of symmetry ( here,
spherical symmetry ), one can use directly the field equations to
match [33-34] ]

 The
wormhole metric is continuous from the throat, $ r = r_0$ to a
finite distance $ r = a $. Now we impose the continuity of $
g_{tt} $ and $ g_{rr}$,

$ {g_{tt}}_{(int)}|_S =  {g_{tt}}_{(ext)}|_S $

$ {g_{rr}}_{(int)}|_S =  {g_{rr}}_{(ext)}|_S $

at $ r= a $ [ i.e.  on the surface S ] since $ g_{\theta\theta} $
and $ g_{\phi\phi}$ are already continuous.

\pagebreak

 The continuity of the metric then gives generally

$ {e^{\nu}}_{int}(a) = {e^{\nu}}_{ext}(a) $ and $
{e^{\mu}}_{int}(a) = {e^{\mu}}_{ext}(a) $.

Hence one can find

\begin{equation}e^{\nu}= ( 1 - \frac{2GM}{a}) \end{equation}

and $  1 - \frac{b(a)}{a} = ( 1 - \frac{2GM}{a})  $ i.e. $ b(a) =
2GM $

This implies $ \frac{D}{a} = 2GM $

Hence,

\begin{equation} a = \frac{D}{2GM} \end{equation}

i.e. matching occurs at $  a = \frac{D}{2GM} $.

The interior metric $ r_0 < r \leq a $ is given by

\begin{equation}
               ds^2=  - [ 1-\frac{D}{a^2}]dt^2+ \frac{dr^2}{[ 1-\frac{D}{r^2}]}+r^2( d\theta^2+sin^2\theta
               d\phi^2)
          \end{equation}

The exterior metric $ a \leq r < \infty   $ is given by

\begin{equation}
               ds^2=  - [ 1-\frac{D}{ar}]dt^2+ \frac{dr^2}{[ 1-\frac{D}{ar}]}+r^2( d\theta^2+sin^2\theta
               d\phi^2)
          \end{equation}

Since wormhole is not a black hole, so we have to impose the
condition $ a > 2GM $ and mass of the wormhole is given by  $ M =
\frac{D}{2Ga} $.

One can note that the geometry of equation(13) is same as the
Ellis wormhole geometry [35].

\pagebreak

We see that the metric coefficients continuous at the junction
i.e. at S. However, the metric need not be differentiable at the
junction and the affine connection may be discontinuous there.
This statement may be quantified in terms of second fundamental
form of the boundary.

 The second fundamental forms associated with the two
sides of the shell are [26-32]

\begin{equation}K_{ij}^\pm =  - n_\nu^\pm\ [ \frac{\partial^2X_\nu}
{\partial \xi^i\partial \xi^j } +
 \Gamma_{\alpha\beta}^\nu \frac{\partial X^\alpha}{\partial \xi^i}
 \frac{\partial X^\beta}{\partial \xi^j }] |_S \end{equation}

where $ n_\nu^\pm\ $ are the unit normals to $S$,

\begin{equation} n_\nu^\pm =  \pm   | g^{\alpha\beta}\frac{\partial f}{\partial X^\alpha}
 \frac{\partial f}{\partial X^\beta} |^{-\frac{1}{2}} \frac{\partial f}{\partial X^\nu} \end{equation}

with $ n^\mu n_\mu = 1 $.

[  $\xi^i$ are the intrinsic coordinates on the shell with $f =0
$ is the parametric equation of the shell S and $-$  and
 $ + $ corresponds to interior ( Wormhole ) and exterior (
Schwarzschild )]. Since the shell is infinitesimally thin in the
radial direction there is no radial pressure. Using Lanczos
equations [26-32], one can find the surface energy term $\sigma$
and surface tangential pressures $ p_\theta = p_\phi = p $ as

$     \sigma =  - \frac{1}{4\pi a}[ \sqrt{e^{-\mu}}]_-^+ $

$     p =   \frac{1}{8\pi a}[ ( 1 + \frac{a \nu^\prime }{2})
\sqrt{e^{-\mu}}]_-^+ $

The metric functions are continuous on S, then one finds $ \sigma
= 0 $. As $ \nu^\prime_- $ = 0 , one gets,  $     p =
\frac{1}{8\pi a} \frac { \frac{GM}{a}}{\sqrt{1 - \frac{GM}{a}}} $.

Hence one can match interior wormhole solution with an exterior
Schwarzschild solution in the presence of a thin shell. The total
metric of the whole spacetimes is given by the equations (13) and
(14) which are joined smoothly.

 Now we use the following analytical method for strong
field limit gravitational lensing [22]. Consider a light ray from
a source(S) is deflected by the lens (L) of mass M and reaches an
observer(O). The background space time is taken asymptotically
flat, both the source and the observer are placed in the flat
space time. The line joining the lens and the observer(OL) is
taken as the optic axis for this configuration.

\pagebreak

 The
position of the source and the image are reflected through the so
called lens equation [22]

\begin{equation}
               \tan \theta - \tan \beta = d[ \tan\theta + \tan (
               \alpha- \theta )]
          \end{equation}

where $\alpha $ is the deflection angle, $ d =
\frac{d_{LS}}{d_{OS}} $ and $\beta$, $\theta$ are the angular
position of the source and the image with respect to the optic
axis. $d_{LS}$ and $d_{OS}$ distances between observer and source
and lens and source. For a general static and spherically
symmetric space time of the form

               $ds^2=  - A(x)dt^2 + B(x) dx^2 + C(x)( d\theta^2+sin^2\theta
               d\phi^2)$ ,

the deflection angle is given by as function of the closest
approach $ x_0 = \frac{r_0}{2M}$ by $ \alpha (x_0) = I(x_0) - \pi
   $

where
\begin{equation}
               I(x_0) = \int_{x_0}^{\infty}  \frac{2\sqrt{B(x)}dx}
               {\sqrt{C(x)}\sqrt{\frac{C(x)A(x_0)}{C(x_0)A(x)} -
               1}}
          \end{equation}
Now according to Ref.[22],  one has to expand the integral in $
I(x_0) $ around the photon sphere to obtain $ \alpha (x_0)$ and
then, using the lens equation,  one can obtain the position of the
produced images. For $ x_0 \geq a $ i.e. closest approach outside
the wormhole's mouth, then the lensing effect exactly the same as
that produced by Schwarzschild metric [24].

\begin{equation}
               \alpha (x_0) = -2\ln (  \frac{2x_0}{3} - 1 ) -
               0.8056
                         \end{equation}

For closest approach inside the wormhole's mouth i.e. $r_m <x_0
<a$  ( $r_m =  $ throat$ $ radius $ $ ), the deflection angle is
given by

\begin{equation}
               \alpha _e = -2\ln (  \frac{2a}{3} - 1 ) -
               0.8056
                         \end{equation}

Now the contribution of the C-field wormhole metric ( interior
metric ) is given by

\begin{equation}
               I(x_0) = 2 \int_{x_0}^{a}  \frac{dx}
               {\sqrt{x^2 ( 1 - \frac{x_m^2}{x^2} )( \frac{x^2}{x_0^2}  -
               1)}}
          \end{equation}

          where $ x = \frac{r}{2M} $ and $ x_m = \sqrt{D}= $
           $throat$$
           $ radius $ $.

\pagebreak

          The above integral can be written as ( using $ y = \frac{x}{x_0}
                    $ )

\begin{equation}
               I(x_0) = 2 \int_1^{\frac{a}{x_0}}  \frac{dy}
               {\sqrt{y^2 ( 1 - \frac{x_m^2}{x_0^2y^2} )( y^2  -
               1)}}
          \end{equation}

Now expanding around 1 in the square root term up to the second
order in y, we have

\begin{equation}
               I(x_0) = 2 \int_0^{\frac{a}{x_0} - 1}  \frac{dz}
               {\sqrt{Az + B z^2  }} \equiv  2 \int_0^{\frac{a}{x_0} - 1}
               f(z, x_0) dz
          \end{equation}

          where $ z = y -1 $ and $ A = 2( 1 - \frac{D}{x_0^2}) $ , $ B = ( 5 -
          \frac{D}{x_0^2}) $.
If $ A = 0 $, then $ f(z, x_0) $ diverges as $ z \rightarrow 0 $.
In other words, for $ A = 0 $ , one can get the radius of the
photon sphere as $ x_{ps} = x_m = \sqrt{D} $. From the above
integral, one can get the following exact analytic form as

\begin{equation}
               I(x_0) = \frac{2}{\sqrt{5 -
          \frac{D}{x_0^2}}}\ln [  2 \sqrt{{\frac{(5 -
          \frac{D}{x_0^2})^2( \frac{a}{x_0} -1)^2}{4 (1 -
          \frac{D}{x_0^2})^2}} + \frac{(5 -
          \frac{D}{x_0^2})( \frac{a}{x_0} -1)}{2(1 -
          \frac{D}{x_0^2})} } + \frac{(5 -
          \frac{D}{x_0^2})( \frac{a}{x_0} -1)}{2(1 -
          \frac{D}{x_0^2})}+1]
          \end{equation}

Hence the total deflection for C-field wormhole is
\begin{equation}
               \alpha_{WH}(x_0) = \alpha _e + I(x_0)
          \end{equation}

Now, graphically,  we compare the deflection angle of C-field
wormhole with the deflection of Schwarzschild. We assume
wormhole's mouth is at $ a = 2 $ AU $  $ and the throat is at $
x_m = \sqrt{D}= 1   $  AU $  $. One can note that the curve due to
C-field wormhole diverges exactly at the wormhole's throat $  x =
x_m = \sqrt{D} $. Also one can see that the angles coincide
outside the wormhole's mouth in both the cases.

          \pagebreak

\begin{figure}[htbp]
    \centering
        \includegraphics[scale=.8]{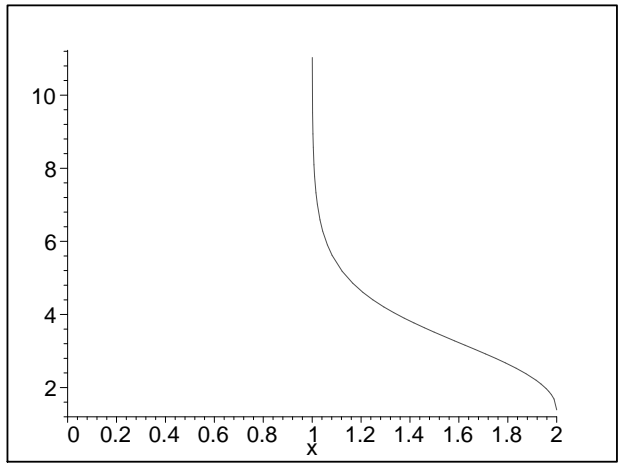}
        \caption{ Deflection angle is a function of closest approach due to the C-field wormhole }
    \label{fig:lens2}
\end{figure}

\begin{figure}[htbp]
    \centering
        \includegraphics[scale=.8]{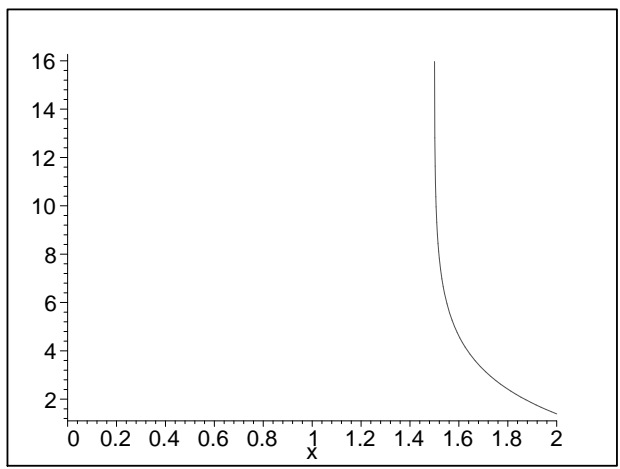}
        \caption{ Schwarzschild deflection angle is a function of closest approach }
    \label{fig:lens1}
\end{figure}

\title{\Huge4. Stability Analysis: }

To study the stability, we match the interior C-field wormhole to
the exterior Schwarzschild solution at a junction interface S,
situated outside the event horizon, $ a > r_b = 2M$, one needs to
use extrinsic curvature of S. The extrinsic curvature or second
fundamental form is defined as $ K_{ij} = n_{\mu ; \nu} e^\mu
_{(i)} e^\nu _{(j)}  $, where $ n^\mu $ is the unit normal 4 -
vector to S and $e^\mu _{(i)}$ are the components of the
holonomic basis vectors tangent to S. In general $K_{ij}$ is not
continuous across S. The discontinuity in the extrinsic curvature
is defined by $ \kappa_{ij} = K_{ij}^+ - K_{ij}^- $. Using the
Darmois-Israel formalism, we write Lanczos equations for the
surface stress energy tensors $S_i^j$ at the junction interface S
as [26-32]
\begin{equation}
               S_i^j = - \frac{1}{8\pi} [ \kappa_i^j - \delta_i^j
               \kappa_k^k ]        \end{equation}

To analyze the dynamics of the wormhole, we permit the radius of
the throat to become a function of time, $ a \mapsto a(\tau)$.

The non trivial components of the extrinsic curvature are given by

\begin{equation}
                K{_\tau^\tau}^+  =  \frac{\frac{M}{a^2} + \ddot{a}}{\sqrt{1-\frac{2M}{a}+
                \dot{a}^2}}
              \end{equation}

\begin{equation}
                K{_\tau^\tau}^-  =  \frac{\ddot{a} - \frac{D\dot{a}}{a(a^2 -D)}  }
                {\sqrt{1-\frac{D}{a^2}+
                \dot{a}^2}}
              \end{equation}

              and

\begin{equation}
                K{_\theta^\theta}^+  =  \frac{1}{a} \sqrt{1-\frac{2M}{a}+
                \dot{a}^2}
              \end{equation}

\begin{equation}
                K{_\theta^\theta}^-  =  \frac{1}{a} \sqrt{1-\frac{D}{a^2}+
                \dot{a}^2}
              \end{equation}

[ we have taken the wormhole space time metric  (10) and
Schwarzschild solution  ]

The surface energy tensor may be written in terms of the surface
energy density $\sigma$ and the surface pressure p as $  S_i^j =
dia( - \sigma, p, p )$.

 Then from Lanczos equations, we
get,

\begin{equation}
                \sigma =  - \frac{1}{4\pi a} [ \sqrt{1-\frac{2M}{a}+
                \dot{a}^2} - \sqrt{1-\frac{D}{a^2}+
                \dot{a}^2}]
              \end{equation}

\begin{equation}
           p  = \frac{1}{8\pi a} [ \frac{1 - \frac{M}{a} + \dot{a}^2 + a \ddot{a}}{
           \sqrt{1-\frac{2M}{a}+ \dot{a}^2}}
            - \frac{1-\frac{D}{a^2}+
                \dot{a}^2 + a\ddot{a} -  \frac{aD\dot{a}^2}{(a^3 -D)}}
                {\sqrt{1-\frac{D}{a^2}+
                \dot{a}^2}}]
              \end{equation}

\pagebreak

 Taking into account conservation identity $
{S^i_j}_{|i} = [ T_{\mu\nu}e^\mu_{(j)}n^\nu ]_-^+ $ and using $
{S^i_\tau}_{|i} = - [ \dot{\sigma} + \frac {2\dot{a}(\sigma +
p)}{a}] $, one gets,

\begin{equation}
                \sigma^\prime =  - \frac{2(\sigma + p)}{ a} + E      \end{equation}

where

 \begin{equation}
              E =  \frac{1}{4\pi a^2} [ \frac{2D}{a(a^2 - D)}]\sqrt{1 - \frac{M}{a} + \dot{a}^2 }
                 \end{equation}

 Rearranging equation (31), one can get the thin shell's
 equation of motion i.e. $
              \dot{a}^2 + V(a)= 0 $
                with the potential is given by

\begin{equation}
              V =  1 + A -B^2
                 \end{equation}

                 where

                 $A = \frac{MD}{a^2C^2} $, $C = 4\pi a^2\sigma $
                 and $B = 2\pi a\sigma + \frac{M + \frac{D}{2a^2}}{4\pi
                 a^2\sigma} $.

Linearizing around a static solution situated at $a_0$, one can
get expand V(a) around $a_0$ to yield

\begin{equation}
              V =  V(a_0) + V^\prime(a_0) ( a - a_0) + \frac{1}{2} V^{\prime\prime}(a_0)
              ( a - a_0)^2 + 0[( a - a_0)^3]
                 \end{equation}

One can verify that at the static solution, $ a = a_0 $, $
V(a_0)= 0 $ and $ V^\prime(a_0)= 0 $. From $ V^\prime(a_0)= 0 $,
one can get an equlibrium relation as $ \frac{dA}{da}\mid_{a_0} =
2BB^\prime\mid_{a_0} $ [ see appendix for exact analytical form ].

 Hence the potential equation reduces to
\begin{equation}
              V =   \frac{1}{2} V^{\prime\prime}(a_0)
              ( a - a_0)^2 + 0[( a - a_0)^3]
                 \end{equation}

The solution is stable iff V(a) has a local minimum at $a_0$.
Then the stability condition is $ V^{\prime\prime}(a_0)> 0 $ i.e.
$ A^{\prime\prime}\mid_{a_0} >  2B^\prime{^2}\mid_{a_0}  +
2B\mid_{a_0}B^{\prime\prime}\mid_{a_0}$.

\pagebreak

 Hence one gets the following expression for stability
condition of the C-field wormhole as

\begin{equation}
             P >  F +R                 \end{equation}

 where P , F and R are given in the appendix.

Also, from (33), one can write,
\begin{equation}
             \eta =   -1 + \frac{a}{2  \sigma^\prime }
             [\frac{2}{a^2 }( p + \sigma) + E^\prime - \sigma^{\prime\prime}]                 \end{equation}

where $\eta = \frac{p^\prime}{\sigma^\prime }$ and this parameter
$\sqrt{\eta}$ is normally interpreted as the speed of sound [28].
Now we shall use $\eta$ as a parametrization of the stable
equilibrium, to show the stability region graphically. To
determine stability region of this solution we use, $ b(a_0) > 2M
$ i.e. when $\sigma < 0$. Now we take $r_0 = \sqrt{D}> 2M, $ so
that junction radius lies outside the event horizon i.e.  $ a_0 >
2M $. The junction radius lies in the following range $2M < a_0 <
\frac{D}{2M}$.

\begin{figure}[htbp]
    \centering
        \includegraphics[scale=.8]{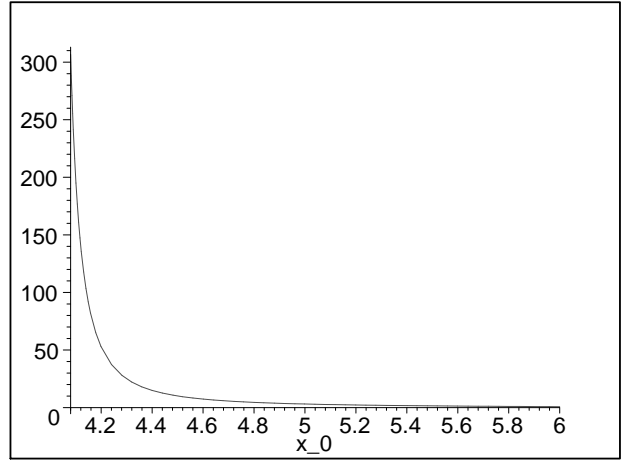}
        \caption{ We define $ x_0 = \frac{a_0 }{M} $ and  choose
        $  \frac{\sqrt{D} }{M} = 3 $. Here we plot $ \eta_{|{(a=a_0)}} $ $ Vs.$ $  x_0 $  for
        the negative surface energy density. The stability region is given
        below the curve.}
    \label{fig:lens3}
\end{figure}

\pagebreak

\title{\Huge5. Conclusion: }

Since the collection of exotic matter is undoubtedly an extremely
hard task, many Scientists are given their attention to the
alternative  theories of gravity to obtain wormhole structure. In
this work, we have considered C-field theory to obtain wormhole
space time. Gravitational lensing seems to be an unique tool, on
a theoretical ground, to detect exotic objects in the Universe,
that, though not yet observed. We have discussed the
gravitational lensing due to a C-field wormhole in the strong
field limit. We have shown that the deflection angle of the
C-field wormhole is very similar to the Schwarzschild solution.
It has been shown that radius of the photon sphere is equal to
the throat of the C-field wormhole. From the graph, one can see
that the deflection angle $\alpha_{WH}$, diverges at the throat.
Since we have matched the interior wormhole solution to the
exterior  Schwarzschild solution across a surface, at $ r = a $,
the deflection angle due to the both the cases should coincide.
The graphs which have been depicted here support this. We have
also analyzed the C-field wormhole solution by matching an
interior solution to exterior  Schwarzschild space time, at a
junction surface. We have explained the stability of the
spherically symmetric shell to linearized perturbation about
static equilibrium solution. We have provided an exact analytical
expression for stability of the C-field wormhole ( see equation
(38)). For a lot of useful information, we have shown the
stability region graphically. We have calculated the range where
the junction radius would lie and have shown that stability
region may be increased by increasing the wormhole throat.

\pagebreak

 { \bf Acknowledgments }

          F.R is thankful to Jadavpur University and DST , Government of India for providing
          financial support. MK has been partially supported by
          UGC,
          Government of India under MRP scheme. We are also grateful to the referee for his valuable comments. \\

\pagebreak

{  \Huge Appendix }
 \centering

$\sigma_0 =  - \frac{1}{4\pi a_0} [ \sqrt{1-\frac{2M}{a_0}} -
\sqrt{1-\frac{D}{a_0^2}}
               ]$

$ \sigma^\prime_0=\frac{1}{4\pi a_0^2}[
\frac{1-\frac{3M}{a_0}}{\sqrt{1-\frac{2M}{a_0}}}
-\frac{1-\frac{2D}{a_0^2}}{\sqrt{1-\frac{D}{a_0^2}}}] $
$
\sigma^{\prime\prime}_0=\frac{1}{4\pi a_0^2}[
\frac{\frac{3M}{a_0^2}}{\sqrt{1-\frac{2M}{a_0}}}
-\frac{M}{a_0^2}\frac{(1-\frac{3M}{a_0})}{(1-\frac{2M}{a_0})^\frac{3}{2}}
-\frac{\frac{4D}{a_0^3}}{\sqrt{1-\frac{D}{a_0^2}}}+
\frac{D}{a_0^3}\frac{(1-\frac{2D}{a_0^2})}{(1-\frac{D}{a_0^2})^\frac{3}{2}}]
- \frac{1}{2\pi a_0^3}[
\frac{(1-\frac{3M}{a_0})}{\sqrt{1-\frac{2M}{a_0}}}
-\frac{(1-\frac{2D}{a_0^2})}{\sqrt{1-\frac{D}{a_0^2}}}] $

$[\frac{dA}{da}]_{a=a_0}= - \frac{MD}{8\pi^2 \sigma_0^2 a_0^4} -
\frac{MD(2 \sigma_0 + a_0 \sigma_0^\prime)}{8\pi^2 \sigma_0^3
a_0^7} $

$ [\frac{dB}{da}]_{a=a_0}=  2 \pi \sigma_0 + 2\pi a_0
\sigma_0^\prime - \frac{D}{8\pi \sigma_0 a_0^5}- (8 \pi \sigma_0
a_0+ 4\pi a_0^2\sigma_0^\prime) \frac{(M +\frac{D}{2a_0})}{(4\pi
\sigma_0 a_0^2)^2} $

$ P=  \frac{\frac{4MD}{a_0^3}}{(4\pi \sigma_0 a_0^2)^2} +
\frac{8MD}{a_0^2}\frac{(8 \pi a_0 \sigma_0 + 4\pi \sigma_0^\prime
a_0^2)}{(4\pi \sigma_0 a_0^2)^3} - \frac{4MD}{a_0}\frac{(8 \pi
\sigma_0 + 16\pi a_0\sigma_0^\prime +4\pi a_0^2
\sigma_0^{\prime\prime})}{(4\pi \sigma_0 a_0^2)^3}+
\frac{12MD}{a_0}\frac{(8 \pi \sigma_0 a_0+ 4\pi
a_0^2\sigma_0^\prime)^2}{(4\pi \sigma_0 a_0^2)^4} $

$ F = 2[2 \pi \sigma_0 + 2\pi a_0 \sigma_0^\prime - \frac{D}{8\pi
\sigma_0 a_0^5}- (8 \pi \sigma_0 a_0+ 4\pi a_0^2\sigma_0^\prime)
\frac{(M +\frac{D}{2a_0})}{(4\pi \sigma_0 a_0^2)^2}]^2 $

$ R =  2[2\pi a_0\sigma_0 + \frac{M + \frac{D}{2a_0^2}}{4\pi
                 a^2\sigma_0}][4 \pi \sigma_0^\prime + 2\pi a_0 \sigma_0^{\prime\prime}
+ \frac{2D}{8\pi \sigma_0 a_0^5}+\frac{D (8 \pi \sigma_0 a_0+ 4\pi
a_0^2\sigma_0^\prime)}{ 2a_0^2(4\pi \sigma_0 a_0^2)^2}-\frac{(M
+\frac{D}{2a_0})}{(4\pi \sigma_0 a_0^2)^2}[8\pi\sigma_0 + 16\pi
a_0 \sigma_0^\prime + 4\pi a_0^2 \sigma_0^{\prime\prime}] +
\frac{2(M +
 \frac{D}{2a_0})}{(4\pi \sigma_0
a_0^2)^3}[8\pi\sigma_0 a_0+4\pi a_0^2 \sigma_0^\prime ]^2] $
\linebreak

The stability condition is $ P > R +F $ i.e. if

 $\frac{\frac{4MD}{a_0^3}}{(4\pi \sigma_0 a_0^2)^2} +
\frac{8MD}{a_0^2}\frac{(8 \pi a_0 \sigma_0 + 4\pi \sigma_0^\prime
a_0^2)}{(4\pi \sigma_0 a_0^2)^3} - \frac{4MD}{a_0}\frac{(8 \pi
\sigma_0 + 16\pi a_0\sigma_0^\prime +4\pi a_0^2
\sigma_0^{\prime\prime})}{(4\pi \sigma_0 a_0^2)^3}+
\frac{12MD}{a_0}\frac{(8 \pi \sigma_0 a_0+ 4\pi
a_0^2\sigma_0^\prime)^2}{(4\pi \sigma_0 a_0^2)^4}> 2[2 \pi
\sigma_0 + 2\pi a_0 \sigma_0^\prime - \frac{D}{8\pi \sigma_0
a_0^5}- (8 \pi \sigma_0 a_0+ 4\pi a_0^2\sigma_0^\prime) \frac{(M
+\frac{D}{2a_0})}{(4\pi \sigma_0 a_0^2)^2}]^2+2[2\pi a_0\sigma_0
+ \frac{M + \frac{D}{2a_0^2}}{4\pi a^2\sigma_0}][4 \pi
\sigma_0^\prime + 2\pi a_0 \sigma_0^{\prime\prime} +
\frac{2D}{8\pi \sigma_0 a_0^5}+\frac{D (8 \pi \sigma_0 a_0+ 4\pi
a_0^2\sigma_0^\prime)}{ 2a_0^2(4\pi \sigma_0 a_0^2)^2}-\frac{(M
+\frac{D}{2a_0})}{(4\pi \sigma_0 a_0^2)^2}[8\pi\sigma_0 + 16\pi
a_0 \sigma_0^\prime + 4\pi a_0^2 \sigma_0^{\prime\prime}] +
\frac{2(M +
 \frac{D}{2a_0})}{(4\pi \sigma_0
a_0^2)^3}[8\pi\sigma_0 a_0+4\pi a_0^2 \sigma_0^\prime ]^2]$

 The equilibrium relation is $\frac{dA}{da}\mid_{a_0} - 2BB^\prime\mid_{a_0}=0$ i.e. if

 $ 2( 2\pi
a_0\sigma_0 + \frac{M + \frac{D}{2a_0^2}}{4\pi
                 a_0^2\sigma_0})( 2 \pi \sigma_0 + 2\pi a_0 \sigma_0^\prime
- \frac{D}{8\pi \sigma_0 a_0^5}- (8 \pi \sigma_0 a_0+ 4\pi
a_0^2\sigma_0^\prime) \frac{(M +\frac{D}{2a_0})}{(4\pi \sigma_0
a_0^2)^2})+  \frac{MD}{8\pi^2 \sigma _0^2 a_0^4} + \frac{MD(2
\sigma_0 + a_0 \sigma_0^\prime)}{8\pi^2 \sigma_0^3 a_0^7}=0$

The values of  $\sigma_0$, $ \sigma^\prime_0$  and  $
\sigma^{\prime\prime}_0$  are given above.


\end{document}